\documentclass[
 reprint,
showpacs,preprintnumbers,
 amsmath,amssymb,
 aps,
 prd,
]{revtex4-1}

\usepackage{graphicx}
\usepackage{dcolumn}
\usepackage{bm}

\begin{document}

\preprint{FERMILAB-PUB-12-254-AE}

\title{On the anomalous afterglow seen in a chameleon afterglow search}

\author{Jason H. Steffen}
 \email{jsteffen@fnal.gov}
\author{Alan Baumbaugh}%
\author{Aaron Chou}%
\author{Ray Tomlin}%
\affiliation{%
Fermilab Center for Particle Astrophysics, P.O. Box 500, Batavia, IL 60510
}%

\author{Amol Upadhye}
\affiliation{Argonne National Laboratory, 9700 S. Cass
Ave., Lemont, IL 60439}%

\date{\today}

\begin{abstract}
We present data from our investigation of the anomalous orange-colored afterglow that was seen in the GammeV Chameleon Afterglow Search (CHASE).  These data includes information about the broad band color of the observed glow, the relationship between the glow and the temperature of the apparatus, and other data taken prior to and during the science operations of CHASE.  While differing in several details, the generic properties of the afterglow from CHASE are similar to luminescence seen in some vacuum compounds.  Contamination from this, or similar, luminescent signatures will likely impact the design of implementation of future experiments involving single photon detectors and high intensity light sources in a cryogenic environment.
\end{abstract}

\pacs{06.60.Ei, 07.20.Mc, 07.60.Dq, 42.15.Eq}
\maketitle


\section{Introduction}

The GammeV Chameleon Afterglow Search (CHASE) reported seeing an anomalous afterglow in their apparatus after having shone a high-power pulsed laser into the bore of a cryogenic vacuum chamber immersed in a magnetic field \citep{Steffen:2010}.  The experiment was designed to produce a population of chameleon particles (scalar or pseudoscalar particles with possible couplings to matter and electromagnetic fields) via induced photon-chameleon oscillations within the magnetic field.  After turning off the laser, this chameleon population would diminish as individual particles re-convert into photons and escape the apparatus \citep{Upadhye:2010}.  This afterglow of photons would indicate the presence of chameleon particles, provided that the properties of the regenerated photons matched the predictions from the chameleon field theory.

The fact that an afterglow signal was observed (hereafter called the ``orange glow'' for reasons described shortly) was troubling as various properties of the orange glow were not consistent with theoretical predictions.  Chameleon theory predicts an afterglow with specific dependence on the laser polarization and the magnetic field strength, as well as equivalent ingoing and outgoing photon energies.  By contrast, the orange glow had none of these properties, being independent of both laser polarization and magnetic field, having outgoing photons at a variety of wavelengths dominated by contributions in the orange and red portions of the spectrum (hence the name), and having an unanticipated dependence on the temperature of the vacuum chamber.

In this article we present all of our data that pertains materially to the characterization of the orange glow signal.  We do not claim any specific explanation of the source or cause of the orange glow, though the dependence upon temperature suggests strongly that the effect is due to some chemical or material property that is excited by the input laser.  Indeed, some of the general properties of the orange glow match luminescent behavior seen in some vacuum products \cite{Cooke:1996}, though the details differ.  The ultimate sensitivity of the CHASE experiment was not limited by the presence of the orange glow, though some modifications to the data analysis software and to the science operations were needed in order to achieve the design goals.  The data and discussion presented here may be useful for the design of future experiments that use high intensity light sources in conjunction with single photon detectors in a cryogenic environment.

This paper is organized as follows.  We discuss the design of the CHASE apparatus (\S \ref{design}) followed by multiple sections presenting data used to characterize the orange glow including some of our initial observations (\S \ref{oddruns}), data taken using broadband optical filters (\S \ref{color}), our initial characterization data and science data (\S \ref{sciruns}), and different vacuum chamber temperatures (\S \ref{tempruns}).  
All of this information is presented in roughly the order that it was gathered.  We discuss potential implications of the orange glow and a comparison with previously observed luminescence from \cite{Cooke:1996} in the discussion (\S \ref{discussion}).

\section{Apparatus design\label{design}}

\begin{figure}
\includegraphics[width=0.45\textwidth]{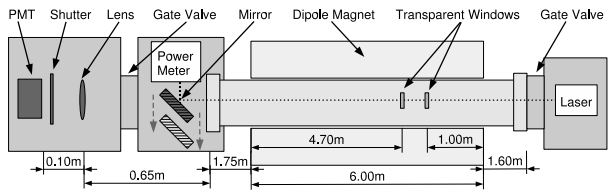}
\caption{Schematic of the CHASE aparatus showing all essential optical elements.}
\label{schematic}
\end{figure}

The CHASE apparatus comprises a laser/optical system, a superconducting magnet, vacuum system, and a single photon detector.  A schematic of the apparatus is shown in Figure \ref{schematic}.  The laser is a Continuum Surelite I-20 Nd:YAG laser, frequency doubled to 532nm.  It provides approximately 3 Watts of green light in 5ns pulses at a rate of 20 Hz.  The laser was shone into the bore of a superconducting Tevatron dipole magnet.  The laser light passed through two anti-reflection coated 1'' BK7 glass windows held within the bore of the magnet by two brass clips attached to an aluminum support rod.  After passing through the magnet, the beam was deflected by a ``pick-off mirror''  into a power meter used to monitor the laser performance.

The magnet itself was energized with fields up to 5 Tesla (currents up to 5040 Amps) and as low as 0.05 Tesla for science operations.  Several calibration runs were conducted with no magnetic field.  When operating, the $\sim 5$cm diameter bore of the magnet cools to 4 Kelvin from the liquid Helium circulating near the magnet coils (within the magnet itself the bore has a slightly square-shaped cross section, but that is of little consequence here).  Cryo-pumping by the cold walls of the chamber in conjunction with three ion pumps yielded a vacuum pressure at or below $10^{-9}$ Torr.  When operating at room temperature the pressure maintained by the ion pumps would occasionally rise slowly by a factor of a few but remained below $10^{-8}$ Torr.

The photodetector was a Hamamatsu H7422P-40 photo multiplier tube (PMT) that had $\sim 30$ Hz of dark rate noise.  The PMT was housed inside a separate, dark ``PMT box'' and was optically isolated from the laser, magnet bore, and power meter by a gate valve.  During operations, the edges of the PMT box were taped closed with metal tape and a dark shroud was laid over the box in order to prevent stray light from the environment from contaminating the measurements.  Ultimately, when the PMT was exposed to the entire apparatus---up to, but not including the ``laser box'' that housed the laser and steering optics---there was a $\sim 1$Hz excess of photons due in part to electromagnetic discharge from the ion pumps \citep{Steffen:2010}.

A typical science run consisted of shining the laser through the bore of the magnet and onto the power meter while the PMT was isolated by the ``PMT gate valve''.  Following laser operations, a second ``laser gate valve'' located near the laser box was closed, the PMT gate valve was opened, and the pick-off mirror was retracted from the beam path by a pneumatic piston.  Any photons streaming from the magnet bore and vacuum system were focused onto the PMT photocathode by a lens with a 4'' focal length.  A custom-made aluminum shutter located between the lens and the PMT was cycled open and closed at intervals of a few seconds (nominally 15 seconds open and 15 seconds closed, though these times were adjusted frequently during the investigation of the orange glow).  These optical components are shown in Figure \ref{schematic}.

A final note on the apparatus was the presence of four gate valves.  The two optical gate valves (PMT and laser) have been mentioned and were used essentially as optical elements to independently isolate the laser and PMT from each other and from the bore of the magnet.  Two other ``vacuum gate valves'' were located on either side of the magnet bore and were used to isolate the magnet bore from the rest of the vacuum system when needed.  Two Residual Gas Analyzers (RGAs) were connected to the apparatus, but were not powered during any of the operations studied herein.  These vacuum gate valves, RGAs, and pressure gauges are not shown in Figure \ref{schematic}.

\section{Initial observation and tests\label{oddruns}}

The orange-colored afterglow was first seen during a test run of the full apparatus.  The magnet was cold but not energized, and all other components were configured for science operations.  Immediately after observing the glow we ran a series of tests to diagnose its source and cause.  Checking the seals and materials inside the PMT box and conducting several test runs using the laser and manipulating the different vacuum gate valves showed that the orange glow originated within the bore of the magnet.

We ran a series of tests to determine how much illumination was required in order to saturate the orange glow signal---including several runs with increasingly shortened laser operations and some where the Q-switch timing on the laser was changed in order to reduce its output power.  Visual inspection of the PMT output showed no noticeable difference between operating the laser for a full 15 minutes and operating it for as little as 15 seconds (roughly the limit of the operation software).  Adjusting the Q-switch until the average power was reduced to 0.2 Watts with an exposure time of 100 seconds finally produced a noticeable drop in the orange glow amplitude.

\begin{table}
\caption{List of relevant parts.}
\label{parts}
\begin{tabular}{lll}
Element & Vendor & Part Number \\ \hline
Laser & Continuum & Surelite I-20  \\
Vacuum windows & Lesker & VPZL-450LYAG  \\
Internal windows & CVI & W2-PW-1025-C-532-0  \\
10nm green filter & Thorlabs & FB530-10  \\
40nm color filters & Thorlabs & FKB-VIS-40  \\
Short-pass filter & Thorlabs & FES0550 
\end{tabular}
\end{table}

We ran a series of tests of the different materials that may have had some residue inside the magnet bore, as well as some spare optical elements.  These elements as well as some of the optical filters that we used to diagnose the color content are shown in Table \ref{parts}.  Placing either an additional interior window (identical to the 1" windows located inside the magnet) or an additional vacuum window in the beam path inside the PMT box showed no increase in the orange glow.  Had the orange glow come from the windows we would have expected a $\sim 50\%$ increase in the signal.

Samples of Kimwipes, Apiezon-L o-ring grease, Ball Vac Kote 44147, and Leybold HE-175 roughing pump oil---all vacuum materials that are commonly used with the Tevatron magnets---were placed in the PMT box near the power meter and pick-off mirror (but not directly in the beam itself) and illuminated with scattered laser light from the pick-off mirror.  We observed no qualitative increase in the orange glow.  An important note, however, is that the PMT box is at room temperature.  No tests of these vacuum materials or the 1'', interior window were done at temperatures near 4 K.  Nevertheless, the vacuum windows are always at room temperature and can therefore be ruled out as the orange glow source.

Following science operations we noted a significant temperature dependence of the orange glow.  Previous studies of the luminescence of vacuum products (particularly Apiezon products) have shown such temperature dependence \citep{Cooke:1996}.  However, there are some slight differences between the effects reported in \citep{Cooke:1996} and the orange glow which may indicate that the orange glow is either a somewhat different effect, comes from a different material, or that it is produced in a somewhat different manner.  We discuss these differences in the observations below, but first present what we know about the orange glow signal.

\section{Broadband color information\label{color}}

We conducted a series of tests using several optical filters inserted between the focusing lens and the PMT.  The expected chameleon afterglow signal would be at the same 532nm green wavelength of the incident laser.  Our first test used a 10nm-wide filter centered at 530nm, where the chameleon signal was expected.  Following that test we used 40nm-wide filters centered at 550nm, 600nm, 650nm, and 700nm as well as a short-pass filter that nominally transmits visible light with wavelengths shorter than 550nm.  The part numbers for these filters are given in Table \ref{parts}.

\begin{figure}
\includegraphics[width=0.45\textwidth]{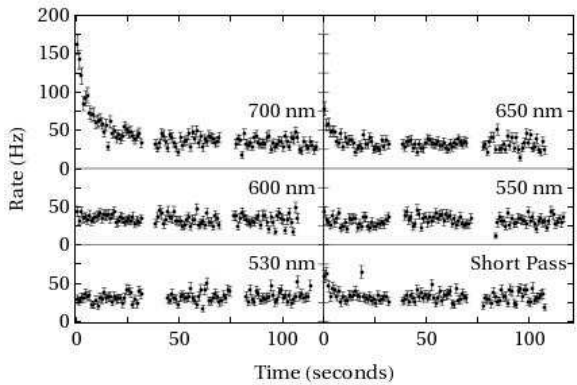}
\caption{Observed afterglow rates for different optical filters listed in Table \ref{parts}.}
\label{filterplot}
\end{figure}

For each color test we operated the laser for three minutes after which afterglow data were aquired for roughly two minutes.  The data reduction and calculation of errors is done following the procedures outlined in \citep{Steffen:2010}.  The resulting time series for each of the six filters is shown in Figure \ref{filterplot}.  The most significant afterglow signals appear in the tests using the 650nm (orange) and 700nm (red) filters.  We saw a 4.1$\sigma$ signal when we used the short-pass filter.  It is not clear from our data if this excess is due to green light that was otherwise filtered by the 530nm and 550nm bandpass filters (e.g. lying in a gap between the wavelength coverage of the two filters), a correlated statistical fluctuation in the PMT hits, a nonlinear upconversion of the incident green light into photons of shorter wavelength, or light bleeding through the filter at longer wavelengths---beyond its effective range\footnote{While the short-pass filter does have a sharp edge in the light that is blocked by the filter, there are always longer wavelengths where the filter is less effective; however, these gaps are typically at infrared wavelengths where the PMT sensitivity is very small.}.

When modeling the time dependence of this signal, a single exponential model yields a good fit to the data from the orange filter.  However, it is not a good fit to the data from the red filter and is a very poor fit to the full-spectrum orange glow data.  Rather, a model of the form
\begin{equation}
\Gamma(t) = \frac{\Gamma_0 e^{-\gamma t}}{1-\xi e^{-\gamma t}} + C
\label{model}
\end{equation}
which gives the decay of a population of excited states through the direct emission of photons and through photon emission mediated by a ``self-interaction'' of two excited particles from within the population.  Here $C$ is the dark rate of the PMT, $\Gamma_0$ normalizes the initial orange glow rate, $\gamma$ is the decay rate for direct emission, and $\xi$ describes the contribution of the self-interaction mediated emission---it is a number between 0 and 1 that describes the relative contribution on the direct and self-interaction components.  This model was used to describe the luminescence observed in \cite{Cooke:1996}), and with different parameter values it gives a reasonable fit to the orange glow data---suggesting that the orange glow may be generated by such a mechanism.

An analysis of the data from the red and orange bands shows a reasonably significant difference in the decay rate $\gamma$ for the two sets of data.  The parameter $\xi$ is not well constrained by the orange-band data that we have.  Nevertheless, a single exponential model for the orange band data (setting $\xi = 0$ in equation (\ref{model})) yields a decay constant of roughly $0.26 \pm 0.06$~s$^{-1}$ while a fit to the red data (allowing $\xi$ to float) gives a decay constant of $0.04 \pm 0.025$~s$^{-1}$, a roughly 3$\sigma$ difference.  A Markov Chain Monte Carlo analysis of the parameter errors in this model, which accounts for its nonlinear nature, yields a similarly small likelihood that these two decay constants are the same.

Unfortunately, the data from the orange band are not adequate to give a meaningful direct comparison of the two colors using equation ($\ref{model}$).  And, with no high resolution spectroscopic data we do not know the full extent of this possible energy dependence of the decay rate.  Nevertheless, this difference in the decay constants is consistent with the observed properties that were reported in \citep{Steffen:2010} which stated that the orange glow had a fast-falling initial phase followed by a slower tail.  The single-exponential model for the orange glow that was used in the CHASE analysis \citep{Steffen:2010} is justified as data from the first 120 seconds were not analyzed.  Thus, the fast decay portion was eliminated \textit{a priori} and, in the limit of late time, equation (\ref{model}) behaves as a pure exponential.  The fitted parameter values of the full-spectrum orange glow that we find in this study (reported below) agree well with the parameter values used in the analysis of CHASE data.



\section{Orange glow characteristics from the CHASE calibration and science data\label{sciruns}}

After finishing the color tests we conducted a series of runs to measure the shape of the orange glow signal and to determine an appropriate cycle for opening and closing the shutter during science operations.  For the science runs, we chose to leave the light unfiltered.  While this added a systematic effect that needed accomodation in the data analysis we decided it was better to have the orange glow information available for study than to remove it via filtering (besides, simply implementing a filter would not guarantee zero transmission of the orange glow and may have necessitated the same adjustments to the data analysis anyway).


We ran a series of 14 ``calibration'' runs with no magnetic field to characterize the temporal profile of the signal and thereby determine the adjustments to our operations and data analysis necessary to reach our design goals; we eventually selected a 15 second open/15 second closed shutter cycle.  Our ``science'' data consisted of 16 runs each with differing laser polarization or a differing magnetic field strength.  The runs at 5 Tesla were repeated for each laser polarization.

\begin{figure}
\includegraphics[width=0.45\textwidth]{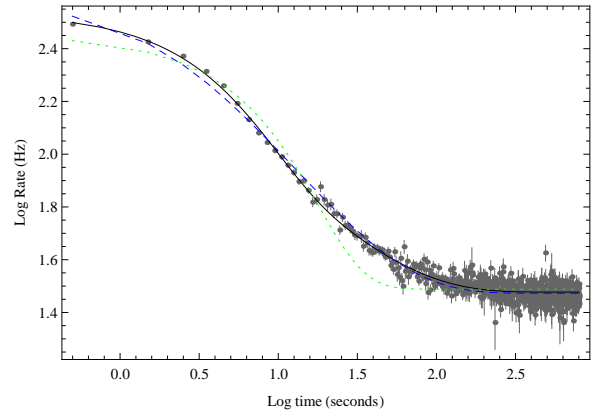}
\caption{Full spectrum orange glow signal using all of the calibration data and the science data.  Shown are the best fit single exponential model (green/dotted), self-interaction model from Equation (\ref{model}) (blue/dashed), and from the preferred model Equation (\ref{goodmodel}) (black/solid).\label{fullplot}}
\end{figure}

\begin{figure}
\includegraphics[width=0.45\textwidth]{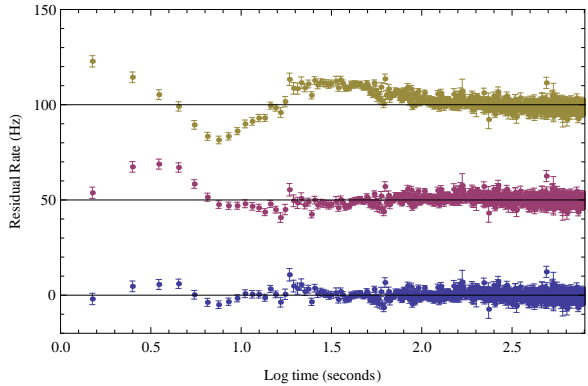}
\caption{Fit residuals for the three orange glow models that we tested.  For clarity, residuals from the the single exponential model (top) and the single self-interaction model (middle) have been shifted vertically by 100 and 50 Hz respectively---and denoted by the horizontal lines.  The preferred model is shown at the bottom.\label{residsplot}}
\end{figure}

Each of these science and calibration runs had at least 90 seconds of illumination by the laser at full power---a time much longer than needed to saturate the orange glow---and four runs illuminated the chamber for five hours.  The amount of time that we took data with the PMT following illumination differed from run to run, but ranged between 6 minutes and 45 minutes.  We will show later that the orange glow was independent of the magnetic field and laser polarization.  For now, we average together the data for all of the calibration and science runs.  Figure \ref{fullplot} shows these data along with three fitted models---a single exponential model, the model given in equation (\ref{model}) from section \ref{color} (used by  \cite{Cooke:1996} in their study), and a ``preferred'' model that incorporates both a quickly falling exponential and a long-term direct plus self-interaction component:
\begin{equation}
\Gamma_{\text{preferred}}(t) = \Gamma_1 e^{-\gamma_1 t} + \frac{\Gamma_2 e^{-\gamma_2 t}}{1-\xi e^{-\gamma_2}} + C
\label{goodmodel}
\end{equation}
where $C$ is the dark rate of the PMT, $\Gamma_1$ and $\Gamma_2$ are the normalizations of the fast and slow components respectively with corresponding decay constants $\gamma_1$ and $\gamma_2$, and $\xi$ characterizes the contribution from the self-interaction of the slow-decay component.  
The residuals from these model fits are shown in Figure \ref{residsplot}.

From Figures \ref{fullplot} and \ref{residsplot} it is clear that a single exponential is not a good fit to the data and that the preferred model is best over the duration of the data.  It also appears that there remains unmodeled physics happening in the first few seconds even with the preferred model.  This unmodeled physics may be from intermediate timescale decays of the population, which would be consistent with an energy dependence of the decay signal as suggested by the color data.  Since the goal here is not to completely characterize the orange glow, but rather to show its basic properties in an effort to guide future experimental design, we do not attempt to model these data further.

\subsection{Polarization and Magnetic Field}

We divided the science runs into groups with vertical and horizontal laser polarization and refit the model from equation (\ref{goodmodel}) in an effort to identify any significant polarization dependence of the orange glow.  We held the dark rate fixed at 30 Hz with these tests because the science runs have gaps in the data that cause poor convergence due to increased parameter degeneracy.  Similarly, we conducted a second test of the science runs by dividing the data into those with magnetic fields less than 1 Tesla and those with magnetic fields greater than or equal to 1 Tesla.  Neither of these tests showed a significant difference between the two respective datasets in the fitted parameter values.  Both tests gave differences that were formally near 1$\sigma$ (estimated from the parameter covariance matrix).  We did not conduct an MCMC error analysis in these cases as the formal errors will over-estimate the differences, implying that these values are conservative.

\section{Variation with temperature\label{tempruns}}

\begin{figure}
\includegraphics[width=0.45\textwidth]{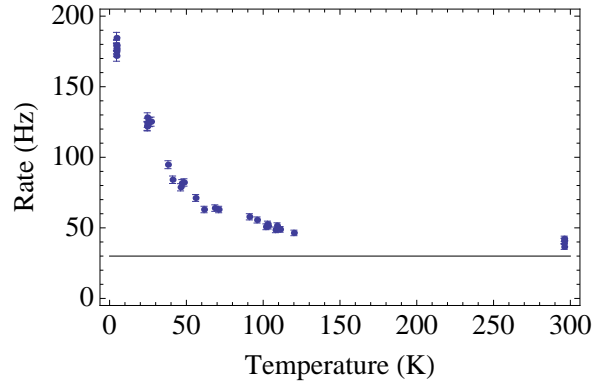}
\caption{Initial PMT rate (10 second average) vs. the input cryogen temperature of the magnet.  The horizontal line indicates the typical PMT dark rate of 30 Hz.\label{ratevstemp}}
\end{figure}

Following science operations we planned a series of runs for further study of the orange glow---but with the magnet at room temperature.  However, the orange glow signal was very weak at best at room temperature.  This temperature dependence was not expected.  Consequently, we flowed cryogens through the magnet with differing input temperatures in order to see how the orange glow changed with temperature.  Figure  
\ref{ratevstemp} shows the initial photon count rate averaged over the first ten seconds as a function of the temperature of the injected cryogen.  As the bore of the magnet approaches liquid helium temperatures the orange glow signal rises significantly.

We note that the magnet, and particularly the magnet operation equipment, is not designed to operate under these conditions and the temperature of the bore of the magnet is not uniform along its length (in some cases changing by several tens of Kelvins.  Thus, while the shape of the temperature dependence seems to be continuous, the orange glow may have a simple, power-law temperature dependence, or it may be caused by some abrupt phase transition.  This transition would occur at different locations within the bore depending upon the input temperature of the cryogen.  Suppose, for example, that the glow is due to some residual gas that boils off the walls during laser operations and then emits light as it re-freezes to the surfaces when the laser is turned off.  Then, only the portion of the bore that is below the freezing point will emit the orange glow and how much of the surface is below that temperature will depend upon the input cryogen temperature.   Given this possibility, we do not know if the temperature dependence of the orange glow is smooth or abrupt.  The CHASE apparatus is not ideal for studying temperature dependence and so we do not provide a more detailed analysis.



\section{Discussion\label{discussion}}

While we do not know the source of the orange glow seen in the CHASE experiment, it does have many properties similar to the observed luminescence from vacuum products reported in \citep{Cooke:1996}.  Specifically, the orange glow shows a temperature dependence, is broadband, lasts for at least a few seconds, and is in the visible portion of the spectrum.  (The temperature dependence, in particular, is likely the reason that the orange glow did not obviously appear when we tested the vacuum products at room temperature if one of those products is indeed the source.)  However, there are some important, though somewhat subtle, differences that distinguish the orange glow from previous reports---though the differences in the setup and execution of the two experiments may explain some of these differences.

First, the wavelength where the orange glow peaks is longer than 670nm instead of between 500 and 600nm as reported in \cite{Cooke:1996}.  It may be that this difference in the peak caused by the fact that our light source is a 532nm wavelength laser instead of the UV lamp and broadband source that were tested in \cite{Cooke:1996}.  This difference in initial illumination could prevent much of the excitation of the material at higher energy leaving only the low frequency tail of the previously reported luminescence.

Second, the orange glow is dominated at later times by a decay with a lifetime that is an order of magnitude longer than that reported in \citep{Cooke:1996}.  It may be that the likely dependence of the decay time with emitted photon energy---a new effect reported here---might account for this difference.  Since the orange glow peaks at lower energy, which also appears to have longer decay times, perhaps the glow seen in \cite{Cooke:1996} is dominated by the higher energy, faster decaying components while the orange glow from CHASE is dominated by the lower energy, longer lifetime excitations that are produced by the lower energy photons that initially illuminate the material.

Finally, the temperature dependence observed in \cite{Cooke:1996} is concave-down instead of concave-up as seen here.  However, the temperature control of the CHASE apparatus is very poor and involves sizeable temperature gradients along the length of the chamber as the operation at intermediate temperatures is far outside of the design of the equipment.  Thus, further investigation of the orange glow is needed in order to better characterize its temperature dependence and to see if the dependence really is different than that reported in \citep{Cooke:1996}.  Here we merely state that the temperature dependence exists and that it is a significant effect.

Further characterization of the orange glow lies beyond the scope of this work.  Nevertheless, the information contained herein is enough to inform the design of future experiments where the detection of single photons is important (e.g. for quantum optics, axion or chameleon searches, etc.).  Either these experiments will need to test the instrument and its associated commercial products that they are using in order to choose build an apparatus that does not exhibit this behavior, or they will need to devise a method of operation to account for this effect.

For CHASE, this orange glow did not preclude the achievement of the design sensitivity---due to the fact that CHASE was looking for either a very long timescale decay, beyond the timescale of the orange glow, or looked for a signal with a specific polarization and magnetic field dependence.  However, if CHASE were to use a photon detector with much less dark rate (such as a Transition Edge Sensor), then it is possible that the orange glow would have proven a fundamental limitation in the absense of using an effective optical or mechanical means to filter the excess photons.




\begin{acknowledgements}
We thank Peter Mazur, Amanda Weltman, and William Wester for their efforts and discussions related to this work the staff at the Fermilab Magnet Test Facility.  This work is supported by the U.S. Department of Energy Office of Science at the Fermi National Accelerator Laboratory Operated by Fermi Research Alliance, LLC under Contract No. De-AC02-07CH11359 and by UChicago Argonne, LLC, operator of Argonne National Laboratory (“Argonne”). Argonne, a U.S.  Department of Energy Office of Science laboratory, is operated under Contract No. DE-AC02-06CH11357. The U.S. Government retains for itself, and others acting on its behalf, a paid-up nonexclusive, irrevocable worldwide license in said article to reproduce, prepare derivative works, distribute copies to the public, and perform publicly and display publicly, by or on behalf of the Government.
\end{acknowledgements}


\begin{thebibliography}{10}
\bibitem{Steffen:2010}
J.H.~Steffen, {\em et al.},
\newblock {\em Phys.Rev.Lett.}, 105, 1803, 2010.
\newblock e-print arXiv:1010.0988.

\bibitem{Upadhye:2010}
A.~Upadhye, J.H.~Steffen, and A.~Weltman.
\newblock {\em Phys.Rev. D}, 81, 015013, 2010.
\newblock e-print arXiv:0911.3906.

\bibitem{Upadhye:2012}
A.~Upadhye, J.H.~Steffen, and A.S.~Chou.
\newblock {\em Phys.Rev. D}, Submitted, 2012.
\newblock e-print arXiv:1204.5476.

\bibitem{Cooke:1996}
D.W.~Cooke and B.L.~Bennett.
\newblock {\em J. of Luminescence}, 63, 283-288, 1996.

\end{thebibliography}

\bibliographystyle{unsrt}

\end{document}